\documentclass[review]{elsarticle}

\usepackage{lineno,hyperref}
\usepackage{color}
\usepackage{verbatim}
\usepackage{graphicx}
\usepackage{epsfig}
\usepackage{amsmath,amssymb}
\usepackage{dsfont}

\modulolinenumbers[5]

\journal{Journal of \LaTeX\ Templates}

\begin{document}

\begin{frontmatter}

\title{SO(3) ``Nuclear Physics'' with ultracold Gases\tnoteref{mytitlenote}}
\tnotetext[mytitlenote]{All authors contributed extensively to carry out the research and writing the manuscript. The authors declare no conflict of interest.}

\author[a]{E. Rico\corref{mycorrespondingauthor}}
\cortext[mycorrespondingauthor]{Corresponding author}
\ead{enrique.rico.ortega@gmail.com}
\author[b]{M. Dalmonte}
\author[c]{P. Zoller}
\author[d,e]{\\D. Banerjee}
\author[d]{M.\ B\"ogli}
\author[d]{P. Stebler}
\author[d]{U.-J. Wiese}

\address[a]{IKERBASQUE, Basque Foundation for Science, Maria Diaz de Haro 3, E-48013 Bilbao, Spain and Department of Physical Chemistry, University of the Basque Country UPV/EHU, Apartado 644, E-48080 Bilbao, Spain}
\address[b]{International Center for Theoretical Physics, 34151 Trieste, Italy}
\address[c]{Institute for Theoretical Physics, Innsbruck University, and Institute for Quantum Optics and Quantum Information of the Austrian Academy of Sciences, A-6020 Innsbruck, Austria}
\address[d]{Albert Einstein Center for Fundamental Physics, Institute for Theoretical Physics, University of Bern, Sidlerstrasse 5, CH-3012 Bern, Switzerland}
\address[e]{NIC, DESY, Platanenallee 6, 15738 Zeuthen, Germany}

\begin{abstract}
An \emph{ab initio} calculation of nuclear physics from Quantum Chromodynamics (QCD), the fundamental $SU(3)$ gauge theory of the strong interaction, remains an outstanding challenge. Here, we discuss the emergence of key elements of nuclear physics using an $SO(3)$ lattice gauge theory as a toy model for QCD. We show that this model is accessible to state-of-the-art quantum simulation experiments with ultracold atoms in an optical lattice. First, we demonstrate that our model shares characteristic many-body features with QCD, such as the spontaneous breakdown of chiral symmetry, its restoration at finite baryon density, as well as the existence of few-body bound states. Then we show that in the one-dimensional case, the dynamics in the gauge invariant sector can be encoded as a spin $S = \frac{3}{2}$ Heisenberg model, i.e., as quantum magnetism, which has a natural realization with bosonic mixtures in optical lattices, and thus sheds light on the connection between non-Abelian gauge theories and quantum magnetism.
\end{abstract}


\end{frontmatter}


\section*{Introduction}

Quantum Chromodynamics (QCD) --- a relativistic non-Abelian $SU(3)$ gauge field theory --- describes the strong interaction between color-charged quarks, anti-quarks, and gluons which are permanently confined inside color-neutral hadrons: mesons consist of quark-anti-quark pairs and a fluctuating number of gluons, while baryons consist of three quarks and a fluctuating number of gluons and quark-anti-quark pairs. At moderate densities, nuclear matter (i.e., the matter inside atomic nuclei or, e.g. in the crust of neutron stars) can be described in terms of nucleons (protons and neutrons, i.e.,\ baryons), whose interaction is dominated by the exchange of pions (i.e.,\ mesons). 

Traditional nuclear physics (formulated in a way which is not directly related to QCD) addresses these dynamics in terms of a many-body Schr\"odinger equation. More recent approaches to nuclear physics use closely related (but more systematic) effective field theory techniques, based on nucleon and pion fields. At very high quark densities (e.g. in the core of neutron stars), on the other hand, one expects nucleons to dissolve into quarks and gluons, thus giving rise to strongly coupled dense quark matter, whose description must necessarily be based on the quantum theory of quark and gluon fields (QCD). Significant progress is being made with the numerical techniques of lattice QCD to understand binding energies in few nucleon systems~\cite{Savage:2016egr}.

The temperature $T$ and the baryon chemical potential $\mu$ (which controls the baryon density in a grand canonical ensemble), define the axes of the QCD phase diagram, which confronts us with the grand challenge to understand the ``condensed matter physics of QCD''~\cite{Rajagopal:2000wf}. While ordinary nuclear matter at moderate densities is reasonably well described on the basis of a many-body Schr\"odinger equation for nucleons, the conjectured phases of color-superconducting high-density quark matter (with or without color-flavor locking) are currently beyond the reach of first principle QCD calculations. 

The real-time dynamics of nuclear matter confronts us with further challenges. While the scattering of light nuclei at moderate energies is, at least to some extent, under control of a many-body Schr\"odinger equation for nucleons, the dynamics of heavy ions colliding at relativistic energies, which leads into the deconfined quark-gluon plasma, is far beyond reach, due to severe sign and complex phase problems of the path integral in real time~\cite{Troyer2005}.

Dense matter systems are associated with greater theoretical challenges, due to the many-body nature of the problem. The condensed matter physics analog of the conserved baryon number $B$ in nuclear physics are the independently conserved numbers of electrons and of the many different species of nuclei. This gives rive to a multi-dimensional phase diagram of ``ordinary'' matter, with a plethora of different phases, ranging from gases, liquids, and solids, to ferro- and antiferromagnets, superfluids, superconductors, and quantum Hall liquids, all the way to spin liquids and high-temperature superconductors. Due to their strong correlations, the latter confront us with similar theoretical challenges as the systems at high quark density. 

In recent years, quantum simulators~\cite{Jaksch2005} have arisen as a promising tool to address hard problems in condensed matter physics. Quantum simulators, whether digital or analog, are engineered cold atomic or ionic systems in an ion-trap or an optical lattice. A large class of interactions can be realized by tuning various parameters of the ion trap, the optical lattice or the species of the trapped particles. The quantum simulators realize Feynman's vision of using quantum degrees of freedom to address some hard problems presented to us by  nature. 

In particular, Hubbard-type models (at non-zero density as well as in real time) have already been quantum simulated with ultracold atoms in optical lattices. Thanks to the high degree of controllability and the novel detection possibilities in these simulators, quantum states involving large-scale entanglement, which cannot be represented classically, have already been realized in the laboratory~\cite{Bloch2012,blatt2012quantum}. A dynamical phase transition has recently been probed in a 51-bit quantum simulator~\cite{51atom}. Quantum simulators have also been used to measure quantities such as entanglement entropies, long thought to belong exclusively to the domain of numerical and analytical studies~\cite{entanglement}. The temperature reached in experiments with ultracold fermionic atoms is low enough to observe long-range antiferromagnetic correlations and to simulate quantum magnetism~\cite{Greif1236362,Mazurenko:2016aa}. 

It is natural to ask whether the hard problems in nuclear and particle physics described above can also benefit from quantum simulation \cite{Cirac2012,Zohar:2015cr,Wiese2014}. Indeed, quantum simulator constructions have already been presented for a variety of Abelian and non-Abelian gauge theories similar to QCD~\cite{Banerjee2013,Tagliacozzo2013,Zohar:2013kb,Stannigel:2014bf,Egusquiza2015}. Recently, the first quantum simulation experiment of an Abelian gauge theory has been performed for the Schwinger model (i.e.,\ QED in $(1+1)$-d) on a 4-site spatial lattice of trapped calcium ions. While a precise implementation of the complex QCD dynamics in ultracold quantum matter is a very ambitious long-term goal, the first quantum simulations of QCD-like theories will aim at some qualitative aspects of the QCD dynamics. For example, quantum simulation experiments of Abelian $U(1)$ gauge theories have been proposed in order to investigate the real-time dynamics of a confining string that connects a pair of external static color-charges~\cite{Martinez:2016oq}. Such a string can break by the creation of dynamical quark-anti-quark pairs. Similarly, the real-time dynamics of a disoriented chiral condensate --- the order parameter for spontaneous chiral symmetry breaking --- can be mimicked in quantum simulations of non-Abelian $U(2)$, $SU(2)$, or even $SU(3)$ gauge theories.

In this paper we consider a lattice gauge model which mimics many of the properties of nuclear physics. We investigate the phase diagram of the model, and show that it shares some fundamental properties with QCD, most prominently confinement, the spontaneous breaking of chiral symmetry, and its restoration at finite baryon density, as well as the existence of few-body bound states, which are characteristic features of nuclear physics. We employ the quantum link model (QLM) formulation of lattice gauge theories~\cite{Horn1981,Orland1990,Brower1999,Chandrasekharan1997}, which works with a finite-dimensional link Hilbert space for the gauge fields. 

The most critical step in implementing a gauge theory on a quantum simulator is to make sure that the gauge symmetry remains intact during the simulation~\cite{Zohar:2015cr,Wiese:2013kk}. Usually, this is guaranteed by implementing an energy penalty, that makes gauge variant violations energetically suppressed, and thus restricts the dynamics to the gauge invariant sub-space~\cite{Wiese:2013kk}. While this strategy is well-suited for Abelian symmetries, non-Abelian ones would generically require fine-tuning since many different, non-commuting constraints must be satisfied at the same time. 

We show how our lattice gauge model, which has a non-Abelian gauge symmetry, can be realised in a quantum simulator platform by encoding the operators directly in the gauge invariant subspace, thus guaranteeing exact gauge invariance. Hence, instead of imposing an energy penalty for gauge variant states, our proposed implementation requires the realization of a nearest neighbor quantum spin Hamiltonian. This encoding strategy is generally applicable  to the whole class of quantum link models, which are extensions of Wilson's formulation of lattice gauge theories. In fact, a similar encoding was crucial to quantum simulate the Schwinger model~\cite{Martinez:2016oq} in the Wilson formulation -- the first quantum simulation of an Abelian lattice gauge theory. However, in the case of the Schwinger model, such an encoding leads to a spin model with long-range interactions~\cite{PhysRevD.56.55}. Quantum link models, on the other hand, permit encoding to local spin Hamiltonians. This not only makes the implementation feasible on different platforms, such as ultracold atoms and molecules trapped in optical lattices, but also establishes a novel connection between non-Abelian lattice gauge theories including matter fields and quantum magnetism~\cite{Lee2006,Lacroix2010}. This further highlights the need to realize experiments for quantum magnetism.

\begin{figure}[t!]
\begin{center}
\includegraphics[width = 1.01\columnwidth]{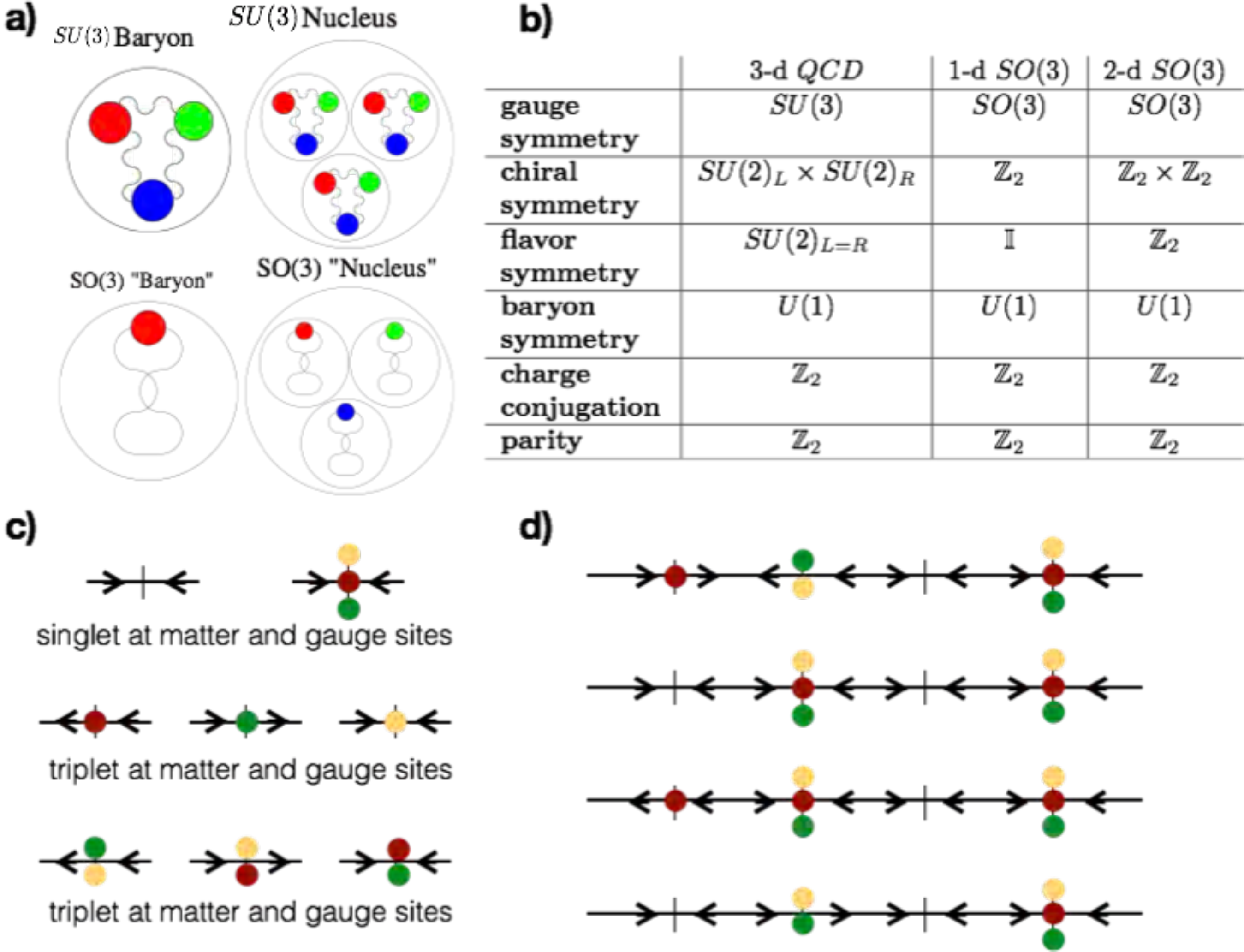}
\caption{ {\it Basic aspects of the $SO(3)$ quantum link model.} {\bf{a)}} Sketches of the different objects in SU(3) and $SO(3)$ gauge theories. In both cases, three-color matter fields are present. Single baryons have different internal structures: in SU(3) gauge theories, they contain three quarks, while in $SO(3)$ theories they can be formed by a single quark paired with a gluon. {\bf{b)}} Summary table of both local and global symmetries in the $(1+1)$-d $SO(3)$ QLM, compared with its $(2+1)$-d counterpart, and with $(3+1)$-d QCD. The model investigated here has the same baryon number symmetry as QCD, and has a non-trivial discrete chiral symmetry (which is simpler than QCD's continuous chiral symmetry). See \ref{symm} for a detailed discussion of the symmetry transformations.  {\bf{c)}} local gauge invariant states in the $SO(3)$ gauge model. Gauss' law implies that the physical subspace contains singlet states in the combined matter and gauge degrees of freedom. {\bf{d)}} Cartoon states for some phases of the $SO(3)$ QLM. From top to bottom: generic configuration with no order; chiral symmetry broken vacuum, where the quark population is arranged in a staggered fashion; "baryon" configuration, where a single baryon is created on top of the vacuum state; "anti-baryon" configuration, where a single "anti-baryon" is created on top of the vacuum state. \label{fig:cartoons} }  
\end{center}
\end{figure}

\section*{Central aspects of nuclear physics}

In order to capture essential aspects of nuclear physics, a toy model non-Abelian gauge theory should at least share some important symmetry features with QCD. First of all, it must necessarily possess a global $U(1)$ baryon number symmetry in order to mimic the corresponding symmetry sectors of QCD. This excludes $U(N) = SU(N) \times U(1)$ gauge theories in which $U(1)$ is gauged and baryons are excluded from the physical spectrum by the
Gauss law. In addition, in order to be at least qualitatively similar to protons and neutrons, the toy model ``baryons'' should be fermions. This excludes theories with an even number of quark colors $N_c$. For example, in $SU(2)$ gauge theory the baryons are bosons and thus have radically different properties than Nature's nucleons.

The conserved baryon number $B$ (which counts the number of quarks minus anti-quarks, divided by the number of quark colors $N_c = 3$) along with isospin $I$ (the number of u-quarks minus d-quarks, divided by the number
of light quark flavors $N_f = 2$) gives rise to a wide variety of strongly interacting systems. These range from the QCD vacuum (at $B = 0, I = 0$) to single protons ($B = 1, I = \frac{1}{2}$) or neutrons ($B = 1, I = - \frac{1}{2}$), light nuclei, such as an $\alpha$-particle (a bound state of two protons and two neutrons with $B = 4, I = 0$), all the way to heavy nuclei or even neutron stars with high baryon density. 

Besides confinement, another essential aspect of nuclear physics is spontaneous chiral symmetry breaking. In QCD with $N_c = 3$ colors and $N_f = 2$ flavors of massless u- and d-quarks the chiral symmetry is $SU(2)_L \times SU(2)_R$. At low temperature and small chemical potential, chiral symmetry is spontaneously broken to the isospin subgroup $SU(2)_{L=R}$, which is manifest in the hadron spectrum. As a consequence of spontaneous chiral symmetry breaking, there are three massless pseudo-scalar Goldstone bosons --- the isospin-triplet of pions -- which mediate the nuclear forces at large distances. In the real world with light (but not exactly massless) u- and d-quarks,
chiral symmetry is, in addition, explicitly broken. The pions then pick up a small mass and turn into pseudo-Goldstone bosons. In the quark-gluon plasma phase at high temperatures or large chemical potential, chiral symmetry is restored in a chiral phase transition (which at non-zero quark mass is washed out to a cross-over).

As any other quantum field theory, QCD must be endowed with a cut-off in order to regularize its ultraviolet divergences. In order to address the nonperturbative strong interaction physics, QCD must be regularized beyond perturbation theory, which is most naturally achieved by replacing continuous space (or space-time) by a discrete lattice. In Wilson's lattice gauge theory~\cite{Wilson74}, the gauge fields are described by group-valued parallel transporter matrices, residing on the links $(x,y)$ connecting nearest-neighbor lattice sites $x$ and $y$, and the matter fields on lattice sites $x$. Monte Carlo simulations of Wilson's lattice QCD are very successful in accurately calculating static properties of hadrons~\cite{Durr:2008yg,Bazavov:2009bb} or thermodynamic properties of the quark-gluon plasma~\cite{Aoki:2006ef,Bazavov:2009bb}. However, they fail already at moderate baryon density or for real-time evolution, due to very severe sign problems. Since Wilson's parallel transporters give rise to an infinite-dimensional link Hilbert space, it is not easy to encode their dynamics in ultracold matter. Quantum link models~\cite{Horn1981,Orland1990,Brower1999,Chandrasekharan1997} provide an attractive alternative for quantum simulations~\cite{Wiese:2013kk}, because they have a finite-dimensional link Hilbert space, while maintaining exact gauge invariance.

\section*{An $SO(3)$ quantum link model for mimicking nuclear physics}
What is the simplest non-Abelian quantum link model that shares the most essential features of nuclear physics --- namely the existence of fermionic baryons with a conserved baryon number, and a spontaneously broken chiral symmetry --- with QCD? Here we propose to investigate an $SO(3)$ non-Abelian quantum link model with ``quarks'' in the adjoint triplet representation. The ``gluons'' also transform in the adjoint representation, which implies that, in this model, a single ``quark'' confined to a ``gluon'' can form a color-neutral fermionic ``baryon'' with a conserved baryon number. In addition, when one uses staggered lattice fermions, the model also has a spontaneously broken discrete chiral symmetry. While the toy model lacks Goldstone pions, and is at best a caricature of QCD, as we will see, it can indeed mimic qualitative aspects of nuclear physics, including the restoration of chiral symmetry at high baryon density or the real-time dynamics of nuclear collisions. In this model, the lowest states in the baryon number $B=2$ and $B=3$ sectors mimic to deuteron and triton, respectively. In order to simplify quantum simulation experiments, we restrict our considerations to $(1+1)$ and $(2+1)$ dimensions.

The quantum link operators that describe the gauge field are denoted by $O_{x,x+\hat{k}}^{ab}$, with color indices $a,b  \in\{1,2,3\}$, site indices $x$, and spatial directions $\hat{k}$. The fermionic matter fields (so-called staggered fermions) $\psi_x^{a}$ transform in the adjoint (triplet) representation. In order to form a colorless "baryon", a single "quark" can bind with a single (color triplet) "gluon", and the resulting "baryon" is a fermion, similar to QCD (see Fig. \ref{fig:cartoons}(c)). This is a significant step beyond the previously considered $SU(2)$ quantum link model~\cite{Banerjee2013}, which only allowed for bosonic "baryons". The Hamiltonian of the model on a $d$-dimensional lattice is given by
\begin{eqnarray}
\label{so3H}
H &=& -t \sum_{x;a,b} \sum_{\hat{k}=1}^{d}  \left[ s_{x,x+\hat{k}} \psi^{a \dagger}_x O^{ab}_{x,x+\hat{k}} \psi^{b}_{x+\hat{k}} + \text{h.c.} \right] + m \sum_{x;a} s_{x} \psi^{a \dagger}_{x} \psi^{a}_x \nonumber \\
&-& \frac{1}{4 g^{2}} \sum_{x;a,b,c,d} \sum_{\hat{k} \neq \hat{l}} O^{ab}_{x,x+\hat{k}} O^{bc}_{x+\hat{k},x+\hat{k}+\hat{l}} O^{cd}_{x+\hat{l},x+\hat{k}+\hat{l}} O^{da}_{x,x+\hat{l}} \\
&+& G \sum_x \left[ \sum_{a} \psi^{a \dagger}_{x} \psi^{a}_x - \frac{3}{2} \right] ^2 + V \sum_{x,\hat{k}} \left[\sum_{a}  \psi^{a \dagger}_{x} \psi^{a}_x - \frac{3}{2} \right]  \left[\sum_{a} \psi^{a \dagger}_{x+\hat{k}} \psi^{a}_{x+\hat{k}} - \frac{3}{2} \right]. \nonumber
\end{eqnarray}
While the first term describes the hopping of the fermions between neighboring sites, mediated by the $SO(3)$ gauge field $O^{ab}_{x, x+\hat{k}}$ on the connecting link, the second term represents the staggered fermion mass, where $s_{x} = \left( -1 \right)^{x_{1}+\cdots + x_{d}}$ and $s_{x,x+\hat{k}}= \left( -1 \right)^{x_{1}+\cdots + x_{k-1}}$. The mass term explicitly breaks the global chiral symmetry and is set to zero in our calculations. The next term describes a plaquette interaction, which is the non-Abelian magnetic field energy contribution to the Hamiltonian. The last two terms describe gauge invariant on-site and nearest-neighbor four-Fermi couplings. Depending on the strength of the couplings $G$ and $V$, a staggered occupation pattern for the fermions is favored, which represents a phase with spontaneously broken chiral symmetry.

As in any gauge theory, the Hamiltonian \eqref{so3H} is invariant under local transformations $V_x = \exp (i \alpha^a_x G^a_x)$, which are represented by unitary transformation in Hilbert space. Here 
\begin{equation}
G^a_x = \psi^\dagger_b T^a_{bc} \psi_c + \sum_{\hat{k}} \left[ L^a_{x,x+\hat{k}} + R^a_{x-\hat{k},x} \right]
\end{equation}
are the infinitesimal generators of gauge transformations, and the $T^a$ are the generators of the so(3) algebra in the adjoint representation, i.e. $T^a_{bc} = -2i \epsilon_{abc}$. Non-Abelian quantum link models have different left and right electric field operators, $L^a_{x,x+\hat{k}}$ and $R^a_{x,x+\hat{k}}$, which play the role of conjugate momenta of the gauge field $O^{ab}_{x,x+\hat{k}}$, and satisfy the canonical commutation relations
\begin{equation}
\label{commutation}
\begin{split}
&\left[ L^a,O^{bd}\right] = 2 i \epsilon_{abc} O^{cd}, ~~\left[ R^a,O^{db}\right] = -2 i O^{dc} \epsilon_{cba}, \\
&\left[ L^a,L^b \right] = 2 i \epsilon_{abc}L^c,~ \left[ R^a,R^b \right] = 2 i \epsilon_{abc} R^c, ~  [L^a,R^b]=0.
\end{split}
\end{equation}
For simplicity, here we have dropped the link indices $(x, x+\hat{k})$; all operators located on different links commute with each other. Physical states $|\Psi\rangle$ must be gauge invariant, i.e., they must obey the Gauss law, $G_x^a|\Psi\rangle = 0$, for all $x$ and $a$.

 For staggered fermions in (1+1)-d, the discrete $\mathbb{Z}_2$ chiral symmetry corresponds to a shift of the link operators and the fermion fields by one lattice spacing, $^S O^{ab}_{x,x+1} = O^{ab}_{x+1,x+2}$, $^S L^{a}_{x,x+1} = L^{a}_{x+1,x+2}$, $^S R^{a}_{x,x+1} = R^{a}_{x+1,x+2}$ and $^S \psi^a_x = \psi^a_{x+1}$. The order parameter for spontaneous chiral symmetry breaking is the chiral condensate
\begin{equation}\label{chicond}
\langle \bar{\psi} \psi \rangle = \left\langle \sum_x (-1)^x \psi^\dagger_x \psi_x \right\rangle.
\end{equation}

\section*{Manifestly $SO(3)$ gauge invariant formulation}

Realising the dynamics of \eqref{so3H} on cold-atom quantum simulators poses two main challenges. First, one has to constrain the dynamics to the gauge invariant subspace. In addition, three- and four-body interactions have to be implemented. The combination of these two requirements makes the realization of non-Abelian lattice gauge theories in cold atomic gases challenging, and only strategies to generate $U(N)$ and $SU(N)$ theories have been devised until now~\cite{Banerjee2013,Tagliacozzo2013,Zohar:2013kb,Stannigel:2014bf,Egusquiza2015}. 

Here we employ a strategy that maintains exact gauge invariance in the experiment without any fine-tuning. As a benefit of the quantum link formulation of $SO(3)$ lattice gauge theories, we can associate spin operators $\sigma^a_{x,+\hat{k}}$ with the left and $\sigma^b_{x+\hat{k},-\hat{k}}$ with the right end of the link $(x,x+\hat{k})$, and identify
\begin{equation}
\label{rep}
\begin{split}
O^{ab}_{x,x+\hat{k}} = &\sigma^{a}_{x,+\hat{k}} \otimes \sigma^{b}_{x+\hat{k},-\hat{k}}, \\
L^{a}_{x,x+\hat{k}} = \sigma^{a}_{x,+\hat{k}} \otimes \mathbb{I}&, ~ \, ~ R^{a}_{x,x+\hat{k}} = \mathbb{I} \otimes \sigma^{a}_{x+\hat{k},-\hat{k}},
\end{split}
\end{equation}
which indeed satisfy \eqref{commutation}. Using this representation \eqref{rep}, it is possible to express the Hamiltonian of the $SO(3)$ QLM in a manifestly gauge invariant way. For this purpose we define the following operators
\begin{equation}
\label{map1}
B_{x,\pm \hat{k}} = \sum_{a} \psi^{a}_{x} \sigma^{a}_{x,\pm \hat{k}}, ~ \, M_{x} =\sum_{a} \psi^{a \dagger}_x \psi^{a}_x, ~ \, \Phi_{x,\pm \hat{k} , \pm \hat{l}} = \sum_{a} \sigma^{a}_{x,\pm \hat{k}} \sigma^{a}_{x,\pm \hat{l}} .
\end{equation}
The total Hilbert space is a direct product of the Hilbert spaces of the fermionic and the spin operators. Because both objects $\psi^{a}_{x}$ and $\sigma^{a}_{x,\pm \hat{k}}$ are color triplets, they can form a color-neutral "baryon" operator. The fermionic part of the Hilbert space can be characterised by the number of "baryons" per site. The operator $B_{x,\pm \hat{k}}$ annihilates a baryon at site $x$, while $B^{\dagger}_{x, \pm \hat{k}}$ creates a baryon. The operator $M_{x}$ is diagonal in an occupation number basis and counts the number of "baryons". Since the baryon number is a conserved quantity, it commutes with the Hamiltonian, $\left[ B = \sum_{x} \left( M_{x} -\frac{3}{2} \right) , H \right]=0$, and the Hamiltonian of the system can be block-diagonalised in each baryon sector individually. The object $\Phi_{x,\pm \hat{k} , \pm \hat{l}}$ can be denoted as a glueball operator. This allows us to write the $SO(3)$ QLM Hamiltonian in terms of the gauge invariant operators
\begin{equation}
\begin{split}
H =& -t \sum_{x,\hat{k}}  \left[ s_{x,x+\hat{k}} B^{\dagger}_{x,+\hat{k}} B_{x+\hat{k},-\hat{k}} + \text{H.c.} \right] + m \sum_{x;a} s_{x} M_x  \\
-& \frac{1}{4 g^{2}} \sum_{x,\hat{k},\hat{l}} \Phi_{x,+\hat{l},+\hat{k}} \Phi_{x+\hat{k},-\hat{k},+\hat{l}} \Phi_{x+\hat{k}+\hat{l},-\hat{l},-\hat{k}} \Phi_{x+\hat{l},+\hat{k},-\hat{l}} \\
+& G \sum_x  M_{x}^2 + V \sum_{x,\hat{k}} M_{x} M_{x+\hat{k}}. 
\end{split}
\end{equation}

\section*{Encoding methods for the $SO(3)$ quantum link model}

\subsection*{Encoding the pure gauge $SO(3)$ quantum link model in $(2+1)$-\textrm{d}}

Here we consider a simple pure non-Abelian quantum link model with an $SO(3)$ gauge symmetry in $(2+1)$-d. The link degrees of freedom are represented by the tensor product of two spin $S = \frac{1}{2}$ subsystems associated with the left and with the right end of the link. The Hamiltonian reduces to plaquette interactions,
\begin{equation}
\begin{split}
H=& \sum_{x;a,b,c,d} \sum_{\hat{k} \neq \hat{l}} O^{ab}_{x,x+\hat{k}} O^{bc}_{x+\hat{k},x+\hat{k}+\hat{l}} O^{cd}_{x+\hat{l},x+\hat{k}+\hat{l}} O^{da}_{x,x+\hat{l}} \\
=& \sum_{x, \hat{k} \neq \hat{l}}  \left(\vec{\sigma}_{x+\hat{k},-\hat{k}} \cdot \vec{\sigma}_{x+\hat{k},+\hat{l}} \right) \left(\vec{\sigma}_{x+\hat{k}+\hat{l},-\hat{l}} \cdot \vec{\sigma}_{x+\hat{k}+\hat{l},-\hat{k}} \right) \\
& \left(\vec{\sigma}_{x+\hat{l},+\hat{k}} \cdot \vec{\sigma}_{x+\hat{l},-\hat{l}} \right) \left(\vec{\sigma}_{x,+\hat{l}} \cdot \vec{\sigma}_{x,+\hat{k}} \right).
\end{split}
\end{equation}

\begin{figure}[t!]
\begin{center}
\includegraphics[width = 0.65\columnwidth]{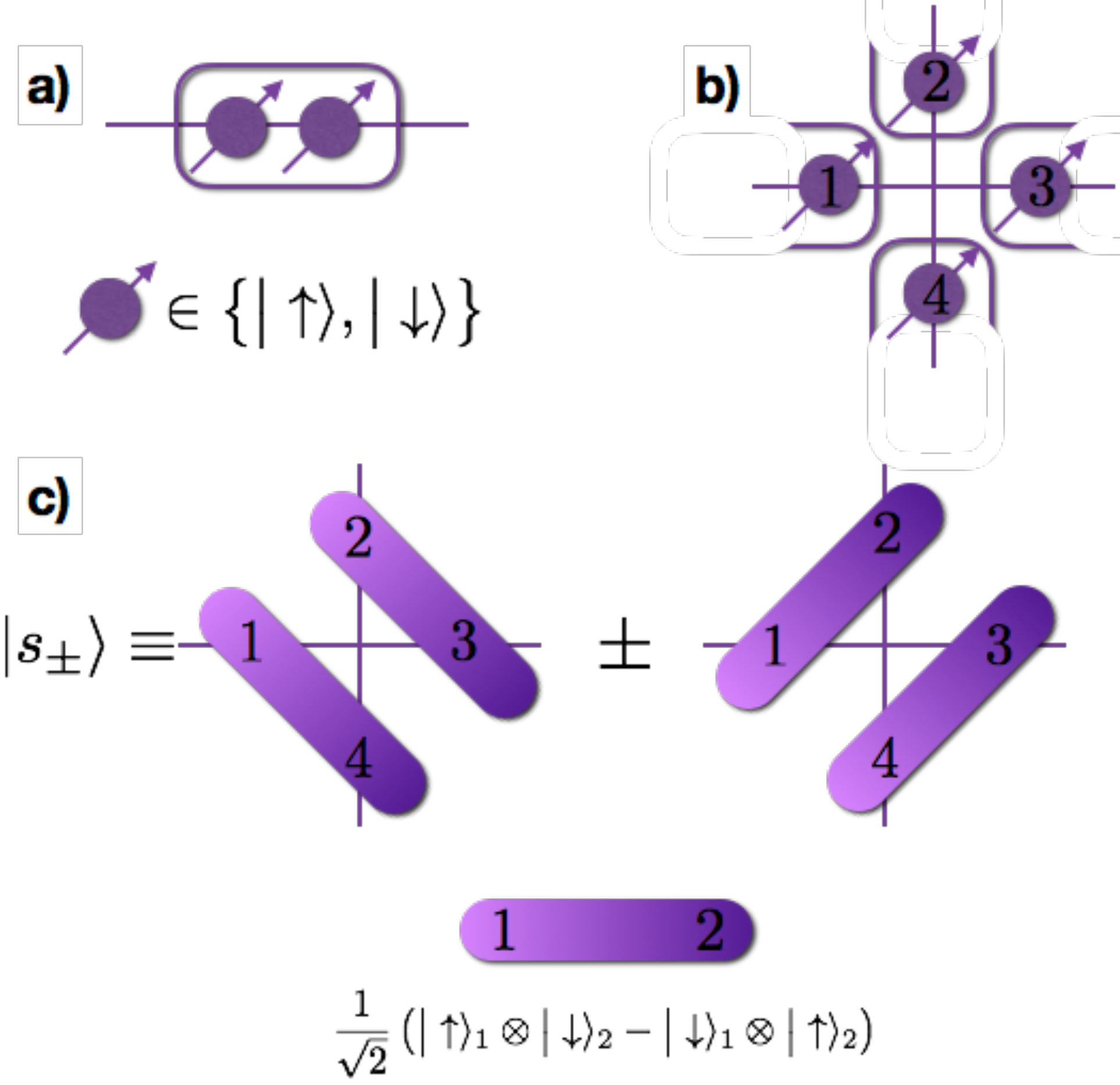}
\caption{{\it Pure $SO(3)$ QLM in} $(2+1)$-d. a) The local Hilbert space of a link described by the tensor product of two spins $S = \frac{1}{2}$. b) Gauss' law implies that the 4 spins $S = \frac{1}{2}$ around each site on a square lattice must form a singlet. c) On the square lattice, the gauge invariant sector with $\vec{G}_{x} | \Psi \rangle =0$ is two-dimensional on the square lattice and it can be represented by the states $|s_{\pm}\rangle$, maintaining the rotational symmetry of the lattice. The gradient in the shading defines an orientation of the singlet which reflects the antisymmetry of the singlet under spin exchange. \label{fig:app}}  
\end{center}
\end{figure}

On the square lattice, each site is touched by 4 links, which brings together 4 spins $S = \frac{1}{2}$, $\vec{\sigma}_{x,\pm \mu}$, which form the gauge generators $\vec{G}_{x} = \sum_{\pm \hat{k}} \vec{\sigma}_{x,\pm \hat{k}}$. Obviously, the Hamiltonian is gauge invariant, i.e., $\left[ \vec{G}_{x} , H \right] = 0$. Gauss' law implies that the 4 spins $S = \frac{1}{2}$ residing around a site must form a singlet. Since $\{2\} \times \{2\} \times \{2\} \times \{2\} = 2~\{1\} + 3~\{3\} + \{ 5\}$, there are two possible ways to form a singlet to obey the Gauss law. As shown in Fig.\ref{fig:app}, invoking rotational symmetry of the square lattice, these two quantum states are defined by
\begin{equation}
\begin{split}
&|s_{-}\rangle = \frac{1}{2} \left( |\uparrow_{1} \downarrow_{3} \rangle - | \downarrow_{1} \uparrow_{3} \rangle  \right) \left( |\uparrow_{2} \downarrow_{4} \rangle - | \downarrow_{2} \uparrow_{4} \rangle  \right) \\
&|s_{+}\rangle = \frac{1}{2 \sqrt{3}} \big(  -2   |\uparrow_{1} \downarrow_{2} \uparrow_{3} \downarrow_{4} \rangle -2  |\downarrow_{1} \uparrow_{2} \downarrow_{3} \uparrow_{4} \rangle \\
&+ |\uparrow_{1} \downarrow_{2} \downarrow_{3} \uparrow_{4} \rangle + |\downarrow_{1} \uparrow_{2} \uparrow_{3} \downarrow_{4} \rangle + |\uparrow_{1} \uparrow_{2} \downarrow_{3} \downarrow_{4} \rangle  + |\downarrow_{1} \downarrow_{2} \uparrow_{3} \uparrow_{4} \rangle \big).
\end{split}
\end{equation}

Projecting onto this subspace, the system is mapped to a vertex model with a two-dimensional local Hilbert space and four-body interactions around every plaquette. Hence, a pure $SO(3)$ QLM in $(2+1)$-d can be mapped onto a problem of quantum magnetism in $(2+1)$-d. Ground state properties, such as the spontaneous breaking of translation invariance or the potential between external charges, are discussed in \ref{spsybr} and \ref{dcon}.

\subsection*{Encoding the $SO(3)$ quantum link model in $(1+1)$-\text{d} with staggered fermions}

Following the same lines as in the previous section, however, now in just one spatial dimension, the physical states must obey the Gauss law, $G^{a}_x |\Psi \rangle = 0$. It turns out that there are only four possible gauge invariant states at each site $x$. Because we are dealing with a model in one spatial dimension, we can describe fermions with spin operators using the Jordan-Wigner transformation. After this, the gauge invariant operators take the form
\begin{equation}
\label{map2}
M_{x} = \begin{pmatrix} 
0 & 0 & 0 & 0 \\
0 & 1 & 0 & 0 \\
0 & 0 & 2 & 0 \\
0 & 0 & 0 & 3 
\end{pmatrix},
B_{x,+} = \begin{pmatrix}
0 & \sqrt{3} & 0 & 0 \\
0 & 0 & 2 & 0 \\
0 & 0 & 0 & \sqrt{3} \\
0 & 0 & 0 & 0
\end{pmatrix},
\end{equation}
with $B_{x,-} = - B_{x,+}$.

The four-dimensional Hilbert space at each site can be identified with the one of a spin $S = \frac{3}{2}$: this is the encoded Hilbert space. After this identification, the baryon number is given by the 3-component of the spin, $M_x=S^3_x+3/2$, and $B_{x,+}^\dagger=S^+_x $ is mapped to the spin raising operators. 

\begin{figure}[t!]
\begin{center}
\includegraphics[width = 1.01\columnwidth]{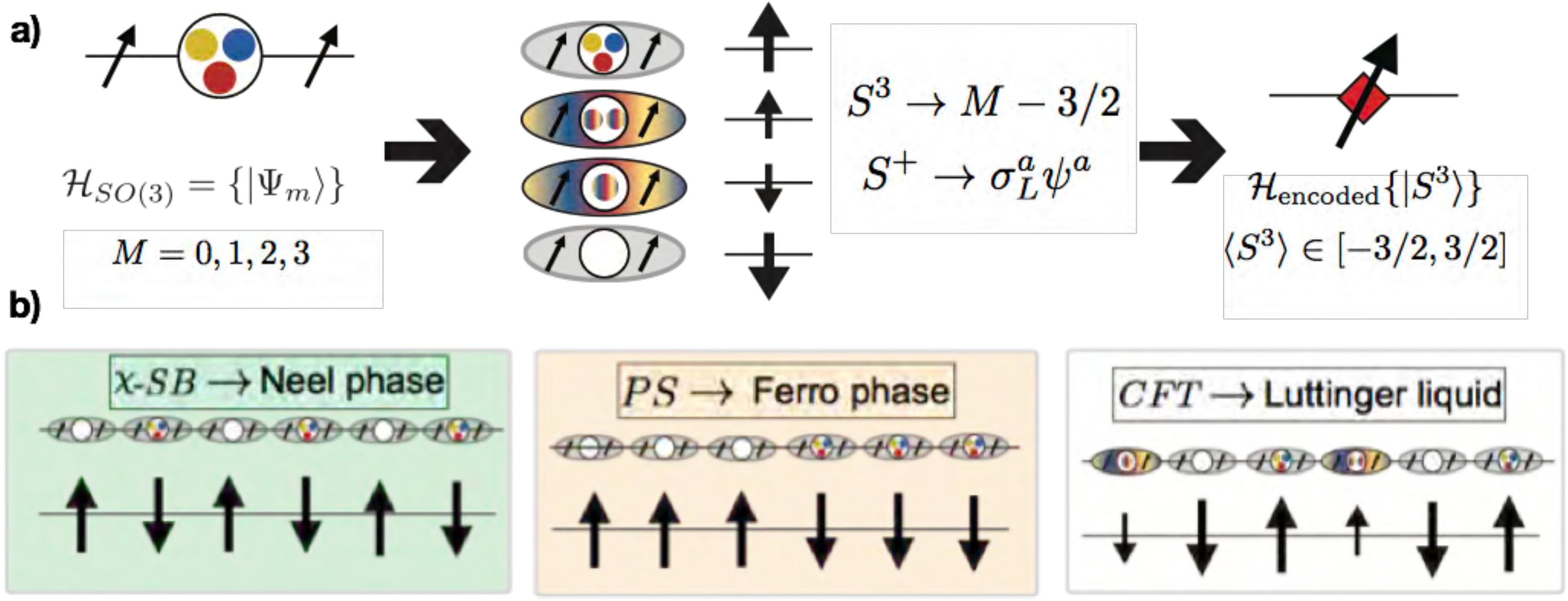}
\caption{ {\it Encoding of the $(1+1)$}-d {\it $SO(3)$ quantum link model.} Upper panel: schematics of the mapping of the on-site Hilbert spaces: in the $SO(3)$ gauge theories, the four local gauge invariant states can be labeled by their baryon number $B$. Each of these states is mapped onto a spin $S = \frac{3}{2}$ state. Lower panel: mapping between the phases of the gauge theory and the spin chain.
\label{fig:enc}}  
\end{center}
\end{figure}

\begin{figure*}[t!]
\begin{center}
\includegraphics[width = 1.01\columnwidth]{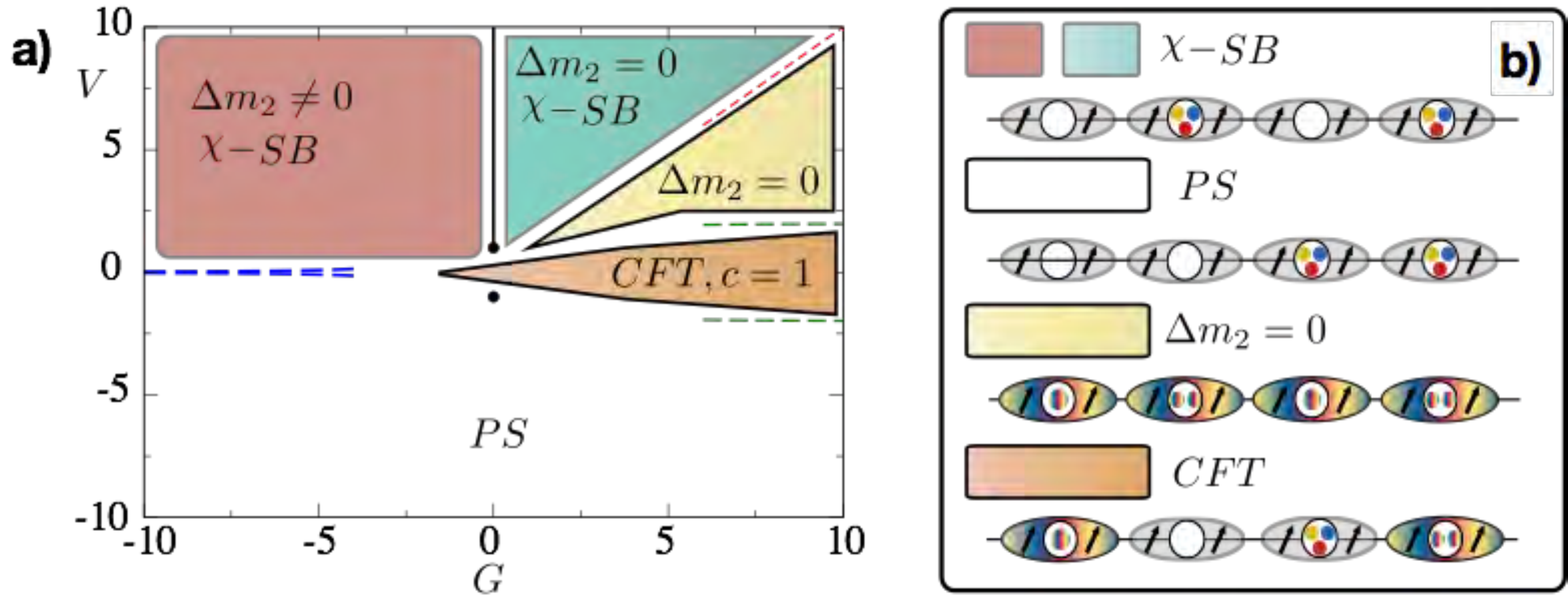}
\caption{ {\it Phase diagram of the $(1+1)$}-d $SO(3)$ {\it QLM.} Panel {\it a)}: Phase diagram in the $G$-$V$ plane in the baryon number $B=0$ vacuum sector. The four shaded areas represent, in clockwise order, a phase with spontaneously broken chiral symmetry ($\chi$SB) (red area) and "nuclear" binding, a phase with symmetry breaking without binding, a gapped phase without chiral symmetry breaking or binding, and a conformal phase (CFT) with central charge $c=1$. The white region represents a phase-separated regime (where binding can occasionally occur as well). The dashed lines are perturbative results for the phase transitions in the various limits; the black, continuous line represents a known confined phase in spin language (see text). Panel {\it b)}: Cartoon states for the strong coupling phases in the different regimes. From top to bottom: chiral symmetry breaking induces a staggered occupation thus breaking translation symmetry by one lattice spacing; in the phase-separated regime, the fermions cluster in a compact spatial region; the intermediate regime between the $\chi$SB and the CFT phase has predominantly doubly and singly occupied sites; in the CFT region, no long-range order is identifiable in the fermion density. 
\label{fig:PD1}}  
\end{center}
\end{figure*}

\begin{figure*}[t!]
\begin{center}
\includegraphics[width = 1.01\columnwidth]{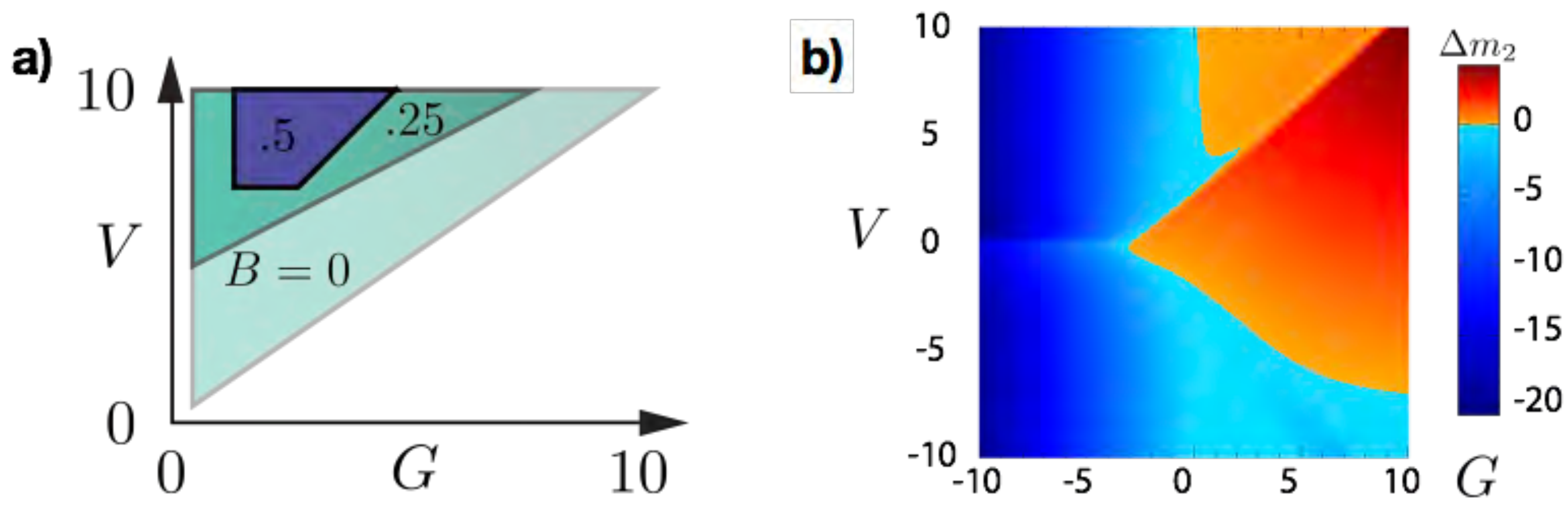}
\caption{ {\it Phase diagram of the $(1+1)$}-d $SO(3)$ {\it QLM.} Panel {\it a)} shows the shrinking of the phase with chiral symmetry breaking as the baryon density $\rho= B/L$ is increased from 0 to $1/(2a)$. Panel {\it b)}: Binding energy $\Delta m_2 = m_2 - 2 m_1$ of a 2-baryon state in the $(1+1)$-d $SO(3)$ QLM on a lattice of size $L/a=10$ with hopping parameter $at=1$. Regions in blue color correspond to a negative binding energy where a 2-baryon bound state exists. In contrast, regions in red color denote regions where binding does not occur. 
\label{fig:PD}}  
\end{center}
\end{figure*}

Thus, up to an irrelevant additive constant, the encoded Hamiltonian is given by
\begin{eqnarray}\label{encodHam}
H_{\textrm{encoded}} & = & -t\sum_{x}(S^+_xS^-_{x+1} + S^+_{x+1}S^-_x) + m \sum_x (-1)^x S^3_x \nonumber \\
&+& G\sum_{x}(S^3_x)^2 + V \sum_{x}S^3_xS^3_{x+1}.
\end{eqnarray}
This mapping establishes an exact correspondence between a non-Abelian lattice gauge theory and a spin $S = \frac{3}{2}$ chain. In particular, the matter-gauge coupling maps to an XY exchange, the staggered mass becomes a staggered magnetic field, and the different four-Fermi couplings transform into single-ion anisotropies ($G$) and nearest-neighbor interactions ($V$). A schematic set-up of the encoding into the spin model is shown in Fig.~\ref{fig:enc}. 

The mapping between quantum link models and spin $S = \frac{3}{2}$ chains sheds light on aspects of "nuclear physics" of this model. Chiral symmetry corresponds to invariance against translation by one lattice spacing, which is indeed spontaneously broken in the N\'eel phase. Two degenerate ground states correspond to the degenerate antiferromagnetically ordered states. Moreover, the bound state scenario also becomes clear in the strong coupling limit $V\gg t$: there perturbations around the N\'eel state make a single spin flip energetically favorable with respect to a double flip when $G$ changes sign. 

Before discussing the complete phase diagram of the model, we note that the idea of encoding Hilbert spaces has already found applications in the context of Wilson's lattice gauge theory. In particular, the Hamiltonian of the $(1+1)$-d QED, the Schwinger model, is known to be exactly mappable to a $S=\frac{1}{2}$ model with long-range interactions~\cite{PhysRevD.56.55}, recently realised in trapped ion experiments~\cite{Martinez:2016oq}. In the context of $SO(N)$ quantum link models, encoding is always possible and leads to local Hamiltonians.  

\section*{Phase Diagram}

Due to their exact correspondence, we immediately gain qualitative and quantitative insights into the phase diagram of the $SO(3)$ model, by mapping its parameters to those of the spin chain. Here we consider $m=0$.

 {\it i)} For $G=0$, the spin $S = \frac{3}{2}$ model supports three phases~\cite{schulz1986phase}. For $V /t>1$, there is a N\'eel antiferromagnetic phase that corresponds to the chirally broken phase in the gauge theory. For $|V/t|>1, V< 0$, there is a ferromagnetic phase that instead translates into a phase-separated regime, in which "baryons" accumulate in a fraction of the volume surrounded by vacuum. Finally, the intermediate regime $-t \lesssim V \lesssim t$ is characterised by a gapless phase with emergent conformal symmetry. 

{\it ii)} In the $G\gg V, t$ limit, the system is effectively mapped onto a spin $S = \frac{1}{2}$ Heisenberg model, with a conformal window $-2t \lesssim V \lesssim 2t$. 

{\it iii)} In the $|G|\gg V,t; G<0$ limit, the dynamics is effectively described by a spin $S = \frac{1}{2}$ Heisenberg model, since the only two states populated at each site have $M_x=\pm3/2$. In this regime, the N\'eel, conformal, and ferromagnetic phases are separated by two critical lines at $V \approx \pm \frac{9t^3}{4G^2}$. 

These predictions are marked by dashed lines in Fig.~\ref{fig:PD1}, and are in very good agreement with the numerical results. In the next section, we will further elaborate on those features of the different phases.

\subsection*{Mimicking nuclear physics: restoration of chiral symmetry}

\begin{figure}[t!]
\begin{center}
\includegraphics[width = 0.6\columnwidth]{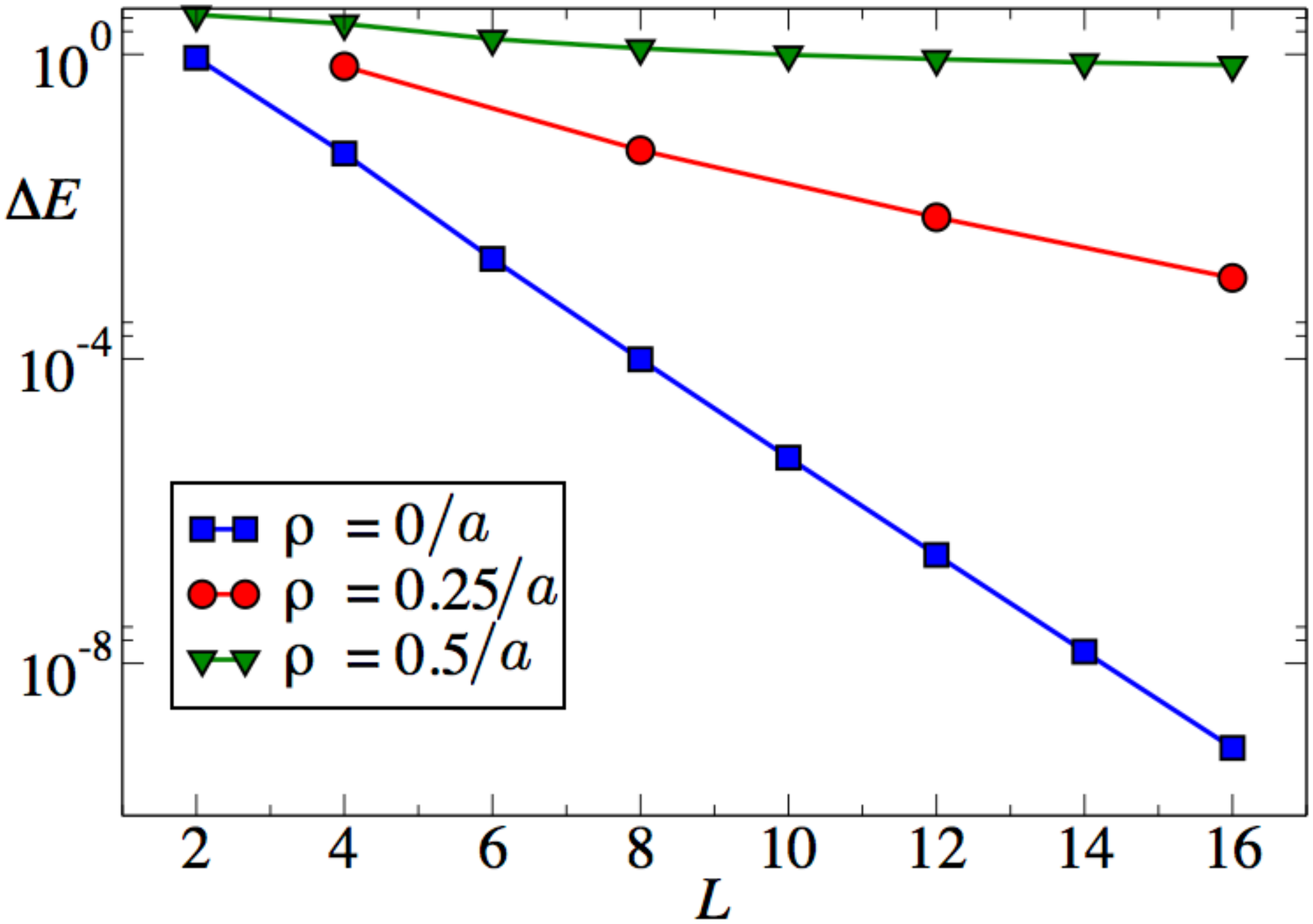}
\caption{{\it Chiral symmetry breaking}. Gap scaling as a function of the system size in the chiral symmetry broken phase. The exponential decay of the energy difference for $G/t = 2$,$V/t = 6$ of the almost degenerate vacua indicates chiral symmetry breaking in the vacuum (i.e. for zero baryon density $\rho = B/L = 0$). Chiral symmetry remains broken at $\rho=1/(4a)$, but is restored for baryon density $\rho = 1/(2a)$. Errors are much smaller than the size of the symbols.}
\label{fig:csb}  
\end{center}
\end{figure}

Just as in QCD, it is interesting to investigate the restoration of chiral symmetry breaking as a function of the baryon density.  First principles lattice QCD calculations are not yet able to tackle this challenging question due to severe sign and complex weight problems. An ultimate goal of our approach is to eventually address these questions using quantum simulators. In order to ensure that the $SO(3)$ toy model indeed shares important features with QCD, and in order to validate future implementations in ultracold atom experiments, we now study the fate of chiral symmetry using exact diagonalization for moderate system sizes by investigating the scaling of the finite-volume energy gap $\Delta E = E^1_{B=0} - E^0_{B=0}$. In a phase with spontaneous chiral symmetry breaking, the energy gap decreases exponentially with the system size, i.e., $\Delta E\sim  \exp (-\alpha L)$. Our results are illustrated in Fig.~\ref{fig:csb} where we show these energy differences for $G/t = 2$, $V/t = 6$. The exponential fall-off is obvious in the vacuum sector, which indeed confirms spontaneous chiral symmetry breaking. As we switch on a finite baryon density $\rho= B/L = 1/4a$ (on a finite lattice with a lattice spacing $a$ and spatial extent $L=Na$), the plot again indicates an exponential fall-off. For $\rho= B/L = 1/2a$, however the exponential decay is lost, indicating restoration of chiral symmetry at this finite baryon density. This study was repeated at different values of the couplings $G/t$ and $V/t$ to determine the phase diagram in Fig.~\ref{fig:PD}.

\begin{figure}[t!]
\begin{center}
\includegraphics[width = 0.65\columnwidth]{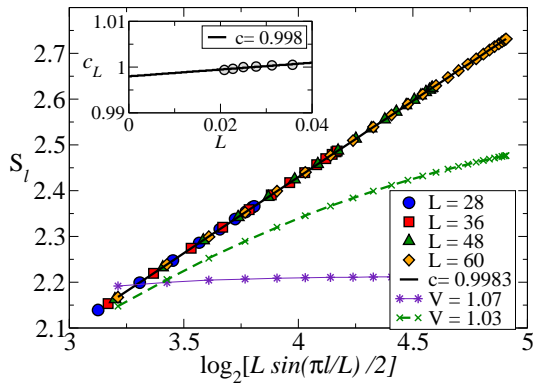}
\caption{{\it Conformal window.} Entanglement entropy for an $SO(3)$ lattice gauge theory as a function of the sub-system length. Symbols: conformal phase ($V=0, G=6t$) for various system sizes. The data collapse on the line with central charge $c=0.9983$ (black line); the finite-size scaling is illustrated in the inset. Thin and dashed lines: two points in the gapped phase, $L=60, G=0, V / t = 1.03, 1.07$ respectively (curves have been shifted by a constant to improve readability). The entropy saturates at small block length, indicating the presence of a finite correlation length. Errors are much smaller than the size of the symbols.}
\label{fig:cft}  
\end{center}
\end{figure}

\subsection*{Mimicking nuclear physics: few-body bound states}

As we have seen, the $SO(3)$ model has a spontaneously broken $\mathbb{Z}_2$ chiral symmetry as well as an unbroken $U(1)$ baryon number symmetry. It therefore mimics certain qualitative features of nuclear physics. Another important characteristics of nuclear physics is the existence of bound states with baryon number $B \geq 2$, i.e.\ stable nuclei. This feature is again shared by the $SO(3)$ model, at least at a qualitative level. Let us first define the mass gap $M_B = E_B^0 - E_{B=0}^0$ as the difference between the energy of the ground state in the baryon number $B$ sector, $E_B^0$, and the vacuum energy $E_{B=0}^0$, in the momentum 0 sector. The baryon mass $M_{B=1}$ is the mass gap in the sector with one baryon. $SO(3)$ ``nuclear'' binding of states with baryon number $B \geq 2$ arises if $M_B < B M_1$. By exact diagonalization of systems up to $N=12$ sites, we have identified a large region in the phase diagram (see Fig.~\ref{fig:PD}b), in which 2-baryon bound states (with $M_2 < 2 M_1$) exist. In different parts of this region, various multi-baryon bound states are also possible. We have verified that the exact diagonalization results (which are limited to moderate volumes) only have small finite-size effects. This shows that the simple $SO(3)$ model indeed mimics {\it nuclear} binding, at least at a qualitative level, and thus shares this important feature with QCD.

\subsection*{Mimicking beyond the standard model physics: conformal window}

Besides the massive chirally broken phase and the chirally restored phase at high baryon density, we also find a massless conformal phase. Interestingly, $SO(3)$ non-Abelian gauge theories with fermions in the adjoint representation have also been discussed as possible extensions of the standard model of particle physics, in order to address the hierarchy problem, i.e., the naturalness of a light Higgs particle~\cite{Hill:2002ap,DelDebbio:2010hx}. In such models, electroweak symmetry breaking is induced by spontaneous chiral symmetry breaking of techni-fermions, whose chiral condensate is naturally small near a conformal window. Hence, our $SO(3)$ model can also mimic certain qualitative features of technicolor theories.
 
The above properties provide information on the phase diagram of the model, and, in particular, for which values of the bare parameters one expects to see QCD-like behavior and related "nuclear physics". In order to characterize it, we have computed the central charge of the underlying conformal field theory (CFT) using density-matrix-renormalization-group (DMRG) simulations~\cite{White1992}. The latter provide direct access to the bipartite entanglement entropy, $S_A^{(l)} = \text{Tr}\rho_A\log \rho_A$, where $\rho_A$ is the density matrix of a given bipartition of size $l$ in a periodic system of size $L$. In conformal phases, the bipartite entanglement entropy is known to scale as $S_A^{(l)} = (c/3)\log (L \sin(\pi l / L)/2)$ up to subleading terms~\cite{vidal2003entanglement,calabrese2009entanglement}, where $c$ is the central charge of the system. In Fig.~\ref{fig:cft}, we present sample results both in the conformal phase, where we found $c=1$ in the entire parameter regime (in agreement with the known result in the $G=0$ case~\cite{hallberg1996critical,dalmonte2012critical}), and in the gapped phases, where instead, the entanglement entropy saturates at a finite value (violet points).   

\section*{Optical lattice realizations}

One of the key advantages of the encoding strategy is that we are able to satisfy exactly the Gauss law constraints by projecting the model to a gauge-invariant subspace. In the case studied here, this is particularly useful, since we have recast an $SO(3)$ lattice gauge model into a local spin system that can be realized in various ways in cold atom systems, with Rydberg atoms or ions. Even more, this encoding strategy allows us to establish a direct connection between quantum magnetism and gauge theories. Below, we discuss a concrete implementation of the encoded spin Hamiltonian using ultracold atoms confined in optical lattices. The starting point is a Bose-Bose mixture in a 1-d lattice, described by the Hamiltonian~\cite{Jaksch2005,Bloch2008} 
\begin{eqnarray}
&H_{\textrm{BB}} = -\sum_{\alpha \in \{\uparrow, \downarrow\}}\sum_x t_\alpha (d^\dagger_{x,\alpha}d_{x+1, \alpha} + 
 d^\dagger_{x+1, \alpha} d_{x,\alpha}) \nonumber\\
&+ \sum_{\alpha \in \{\uparrow, \downarrow\}}\sum_x \frac{U_\alpha}{2}n_{x,\alpha}(n_{x,\alpha}-1) + U_{\uparrow \downarrow}\sum_x n_{x,\uparrow}n_{x,\downarrow},
\end{eqnarray}
where $\alpha=\uparrow, \downarrow$ is the species index, $d^\dagger_{x,\alpha}$ $(d_{x, \alpha})$ are bosonic creation (annihilation) operators that obey the usual commutation relations $[d_{x,\alpha},d^\dagger_{y,\beta}]= \delta_{x,y} \delta_{\alpha,\beta}$ and $n_{x,\alpha} = d^\dagger_{x,\alpha} d_{x,\alpha}$ is the occupation of the corresponding bosonic mode. The first line describes the tunnelling, while the second line describes the intra- and inter-species interactions. The two species can either be two different atomic elements (e.g. $^{87}$Rb-$^{39}$K), two different isotopes ($^{87}$Rb-$^{85}$Rb), or two internal states of the same isotope.

In the total filling case $N=3L$, the strong coupling limit $U_{\uparrow \downarrow},U_\alpha\gg t_\alpha$ constrains the dynamics to a reduced Hilbert space with three particles per site. The local Hilbert space is spanned by the four states of the form $|n_{x,\uparrow}, n_{x,\downarrow}\rangle$, with the constraint $n_{x, \uparrow}+n_{x, \downarrow} = 3$. Using Schwinger bosons as microscopic components ~\cite{auerbach2012interacting}, we define $S = \frac{3}{2}$ operators $S^3_x = ( n_{x, \uparrow} - n_{x, \downarrow} ) / 2$, $S^+_{x} = d_{x,\uparrow}^\dagger d_{x, \downarrow}$. In this way, the effective Hamiltonian dynamics reproduces \eqref{encodHam}, with $t = \frac{2t_\uparrow t_\downarrow}{U}, V = \frac{t_\uparrow^2+t_\downarrow^2}{U} , G = \delta$, where we have defined $U_{\uparrow \downarrow}=U, U_{\uparrow} = U_{\downarrow} = U + \delta$. The staggered mass term can be introduced by using a species-dependent superlattice. 

The parameters $(G, V) $ in the effective Hamiltonian can be tuned independently in two different ways. For species which have a broad intra- or inter-species Feshbach resonance~\cite{Chin:2010fe} (such as mixtures involving $^{39}$K or $^{133}$Cs), the ratios $V/t, G/t$ can be tuned by changing the value of $\delta$. For species where such broad resonances are not available, the fact that the optical lattice confines the atoms in different ways can be exploited to vary the ratio $t_\uparrow/t_\downarrow$, which in turn allows tunability of $t$ and $V$ independently. In both cases, the value of $m$ can be varied by changing the depth of the optical superlattice. Finally, it is important to stress that the baryon number in the original gauge theory $B = \sum_x \left( \psi_x^{a \dagger}\psi_x^a - \frac{3}{2} \right)$ maps onto the total spin imbalance of the Bose-Bose mixture. This quantity can be controlled with a few percent uncertainty in generic mixtures: in the specific case where different Zeeman states are used, it is possible to accurately tune this imbalance using optical pumping techniques, i.e., loading a single Zeeman state onto a deep lattice, and then proceeding to optically pump a certain percentage of atoms into the other Zeeman state.

An important limiting time scale is set by the on-set of three-body recombination. Combinations such as $^{39}$K-$^{87}$Rb should allow for sufficiently long time scales while, at the same time, having sufficiently large interaction strengths -- which can be further increased in 1-d by employing confinement-induced resonances~\cite{Chin:2010fe}.

\subsection*{Observables} Within the present implementation, we discuss how the many-body phase diagram can be experimentally probed. Spontaneous chiral symmetry breaking can be probed by monitoring the chiral condensate order parameter in \eqref{chicond}. In the encoded formulation, this quantity reads $ \langle \bar{\psi} \psi \rangle = \langle \sum_{x} (-1)^x S^z_{x} \rangle$, which in the cold atom realization corresponds to $\langle \sum_x (-1)^x(n_{x,\uparrow}-n_{x,\downarrow}) \rangle$. This quantity can be directly estimated using species-selective single-site imaging, as demonstrated for e.g. in Ref.~\cite{preiss2015quantum}. Assuming that the constraint of three particles per site is fulfilled, it is possible to access the chiral condensate in an even simpler manner. In this case, one has $ \langle \bar{\psi} \psi \rangle =\langle \sum_x (-1)^x (n_{x,\uparrow}- \frac{3}{2}) \rangle$, which amounts to measuring the relative population imbalance between even and odd sites. This quantity is accessible using band-mapping techniques, which do not require single-site imaging. These methods can also be employed to investigate the restoration of chiral symmetry at finite density~\cite{Bloch2008}.

Bound state physics in the original model, and in particular the mass gap, can instead be probed in two different ways. The first one is to extract the mass via the decay of the Green function at long distances, $G(r)=\langle S^+_xS^-_{x+r}\rangle \sim e^{-r m}$. However, this decay might be affected by finite-temperature effects. A more accurate way of probing the low-energy part of the spectrum is by spectroscopic means. As a concrete example, we discuss the measurement of $m_{1}$ for the case of mixtures of different spin states. In this case, the mass gap corresponds to the energy required to change the baryon number by 1: in the spin language, this translates into a magnetization change of $\delta S^3 = 1$, that is, one atom has to change its internal state. This is directly accessible using radio-frequency spectroscopy, which will indicate a sharp resonance peak at the transition frequency $\omega_1 = m_{1}$. The existence of two-particle states will be signalled by an additional peak at $\omega_2 < 2 \omega_1 $, and the corresponding binding energy is $\delta E = \omega_2-2\omega_1 $.

\section*{Conclusions} We have shown how $(1+1)$-d $SO(3)$ lattice gauge theories can be naturally realized using ultracold atoms in optical lattices. The phase diagram of these models features several paradigmatic phenomena, e.g. the presence of stable two-body bound states, phases where chiral symmetry is spontaneously broken, where the broken chiral symmetry is restored at finite baryon density, and emergent conformal invariance. Tailoring our proposal on a specific model has allowed us to explicitly recast its formulation on a spin $S = \frac{3}{2}$ Heisenberg model: this encoding technique has the dramatic advantage of realising gauge invariance exactly, and at the same time bypassing the complex interaction engineering which is required for non-Abelian theories. Our proposal extends considerably the possibility of encoding techniques in quantum simulators, which have recently been experimentally demonstrated for Abelian theories, to non-Abelian models, and shows that quantum simulators are not limited to the $U(N)$ or $SU(N)$ groups discussed so far, but can be extended to $SO(3)=SU(2)$ models as well.

\section*{Acknowledgment} The research leading to these results has received funding from the Schweizerischer Nationalfonds and from the European Research Council under the European UnionÕs Seventh Framework Programme (FP7/2007-2013) ERC grant agreement 339220. Work in Innsbruck is supported by the ERC Synergy Grant UQUAM and the SFB FoQuS (FWF Project No. F4016-N23) E.R. acknowledges support from Spanish MINECO/FEDER FIS2015-69983-P and Basque Government IT986-16. We also acknowledge the EU support from the QuantERA programme under the QTFlag project. 



\appendix

\section{Symmetry transformations}
\label{symm}
In this section we work out the symmetries of the $SO(3)$ quantum link models and study the transformation properties of the various operators.

\begin{itemize}
\item {\bf{Spatial Translations:}} Spatial translations shift the whole system by two lattice spacings in the $k$-direction
\begin{equation*}
\begin{matrix}
 ^{T_{k}}O_{x,y} = O_{x+2\hat{k},y+2\hat{k}}, & ^{T_{k}}\psi_{x} = \psi_{x+2\hat{k}}, \\
 ^{T_{k}}L^{a}_{x,y} = L^{a}_{x+2\hat{k},y+2\hat{k}}, & ^{T_{k}}R^{a}_{x,y} = R^{a}_{x+2\hat{k},y+2\hat{k}}, \\
 ^{T_{k}}M_{x} = M_{x+2 \hat{k}}, & ^{T_{k}}B_{x,\pm l} = B_{x+2 \hat{k},\pm l},  \\
 ^{T_{k}}\sigma^{a}_{x,\pm l} = \sigma^{a}_{x+2 \hat{k},\pm l} .
\end{matrix}
\end{equation*}
This transformation is a symmetry of the Hamiltonian and leads to the conservation of the momentum $p_{k}$. This symmetry can be used to pre-diagonalise the Hamiltonian with respect to the momentum $p_{k}$ and calculate the spectrum for each momentum sector individually.
\item {\bf{Charge Conjugation:}} Besides shifting the system by one lattice spacing in the $k$-direction, the charge conjugation maps particles to anti-particles and performs a specific gauge transformation, which is constant in space and characterized by the gauge transformation matrix
\begin{equation*}
C= \begin{pmatrix}
-1 & 0 & 0 \\
 0 & 1 & 0 \\
 0 & 0 & -1
\end{pmatrix}.
\end{equation*}
In the fundamental representation of $SO(3)$, the operators of the quantum link model transform as
\begin{subequations}
\begin{align}
\nonumber ^{C_{k}}O^{ab}_{x,y} &= O^{ab*}_{x+\hat{k},y+\hat{k}},~~~~~~^{C_{k}}\psi^{a}_{x} = \left( -1\right)^{x_{1} + \cdots + x_{k}}  C^{ab} \psi^{b \dagger}_{x+\hat{k}}, \\
\nonumber ^{C_{k}}L^{a}_{x,y} &= - L^{a*}_{x+\hat{k},y+\hat{k}},~~^{C_{k}}R^{a}_{x,y} = - R^{a*}_{x+\hat{k},y+\hat{k}},\\
\nonumber ^{C_{k}}M_{x} &= 3- M_{x+ \hat{k}},~~~^{C_{k}}B_{x,\pm l} = \left( -1 \right)^{x_{1} + \cdots x_{k}} B_{x+ \hat{k},\pm l}^{\dagger},\\
\nonumber ^{C_{k}}\sigma^{a}_{x,\pm l} &= \sigma^{a*}_{x+ \hat{k},\pm l} .
\end{align}
\end{subequations}
Note that under $C_{k}$ the hopping term in the Hamiltonian transforms into its Hermitian conjugate. The same happens in the continuum for fermions in the adjoint representation. Note further that the quantum link operators $O^{ab}_{x,y}$ and the electric flux operators $L^{a}_{x,y}$ and $R^{a}_{x,y}$ actually undergo an orthogonal transformation. In the fundamental representation with the basis $\left\{ |\uparrow \uparrow \rangle_{x,y} , \, |\uparrow \downarrow \rangle_{x,y} , \,   |\downarrow \uparrow \rangle_{x,y} , \,  |\downarrow \downarrow \rangle_{x,y}    \right\} $, we obtain
\begin{equation*}
\begin{split}
^{C_{k}}O^{ab}_{x,y} &= W_{C} O^{ab}_{x+\hat{k},y+\hat{k}} W_{C}^{\dagger}, \\
^{C_{k}}L^{a}_{x,y} &= W_{C} L^{a}_{x+\hat{k},y+\hat{k}} W_{C}^{\dagger}, \\
^{C_{k}}R^{a}_{x,y} &= W_{C} R^{a}_{x+\hat{k},y+\hat{k}} W_{C}^{\dagger}, 
\end{split}
\end{equation*}
where
\begin{equation*}
W_{C} = \left( i \sigma^{2} \right) \otimes \left( i \sigma^{2} \right) = \begin{pmatrix}
 0 & 0 & 0 & 1 \\
 0 & 0 & -1 & 0 \\
 0 & -1 & 0 & 0 \\
 1 & 0 &  0 & 0
\end{pmatrix}
.
\end{equation*}
This means that the representation of the link operators does not change under charge conjugation, and that the charge conjugation is a symmetry of the Hamiltonian. One can check that the total number of baryons is transformed to $^{C_{k}}B=-B$. This implies that the spectrum is the same for baryon number $B$ and $-B$.
\item {\bf{Parity Transformation:}} The parity transformation inverts all spatial coordinates and interchanges the operators on the left end of a link with the ones on the right end. Therefore the operators transform as
\begin{equation*}
\begin{matrix}
^{P}O^{ab}_{x,y} = O^{ba}_{-y,-x}, & ^{P}\psi^{a}_{x} = \psi^{a}_{-x}, \\
^{P}L^{a}_{x,y} = R^{a}_{-y,-x}, & ^{P}R^{a}_{x,y} = L^{a}_{-y,-x}, \\
^{P}M_{x} = M_{-x}, & ^{P}B_{x,\pm l} = B_{-x,\mp l},  \\
^{P}\sigma^{a}_{x,\pm l} = \sigma^{a}_{-x,\mp l} .
\end{matrix}
\end{equation*}
This transformation is a symmetry of the Hamiltonian. Note that the hopping term transforms into its Hermitian conjugate.
\item {\bf{Chiral Transformation:}} The chiral transformation generates a $\mathbb{Z}_2$ symmetry and corresponds to a shift of the whole system by one lattice spacing in the $k$-direction. Consecutive shifts by one lattice spacing in an even number of directions correspond to a flavor transformation. The translation by one lattice spacing is defined as
\begin{equation*}
\begin{matrix}
^{S_{k}}O^{ab}_{x,y} = O^{ab}_{x+\hat{k},y+\hat{k}}, & ^{S_{k}}\psi^{a}_{x} = \left( -1\right)^{x_{k+1} + \cdots + x_{d}}   \psi^{a}_{x+\hat{k}}, \\
^{S_{k}}L^{a}_{x,y} =  L^{a}_{x+\hat{k},y+\hat{k}}, & ^{S_{k}}R^{a}_{x,y} =  R^{a}_{x+\hat{k},y+\hat{k}}, \\
^{S_{k}}M_{x} = M_{x+ \hat{k}}, & ^{S_{k}}B_{x,\pm l} = \left( -1 \right)^{x_{k+1} + \cdots x_{d}} B_{x+ \hat{k},\pm l},  \\
^{S_{k}}\sigma^{a}_{x,\pm l} = \sigma^{a}_{x+ \hat{k},\pm l} .
\end{matrix}
\end{equation*}
For a non-zero mass $m$, the chiral symmetry is explicitly broken. The flavor symmetry is a symmetry of the Hamiltonian also for $m \neq 0$. 
\end{itemize}

\section{Spontaneous breaking of translation invariance}
\label{spsybr}

Having discussed the encoding of the pure gauge theory with cold atoms, it is interesting to outline some important features of the gauge theory that can be studied. In particular, these highlight the importance of experimental realizations of such gauge theories in cold atom experiments to realise new exotic states. 

\begin{figure}[t!]
\begin{center}
\includegraphics[width = 0.65\columnwidth]{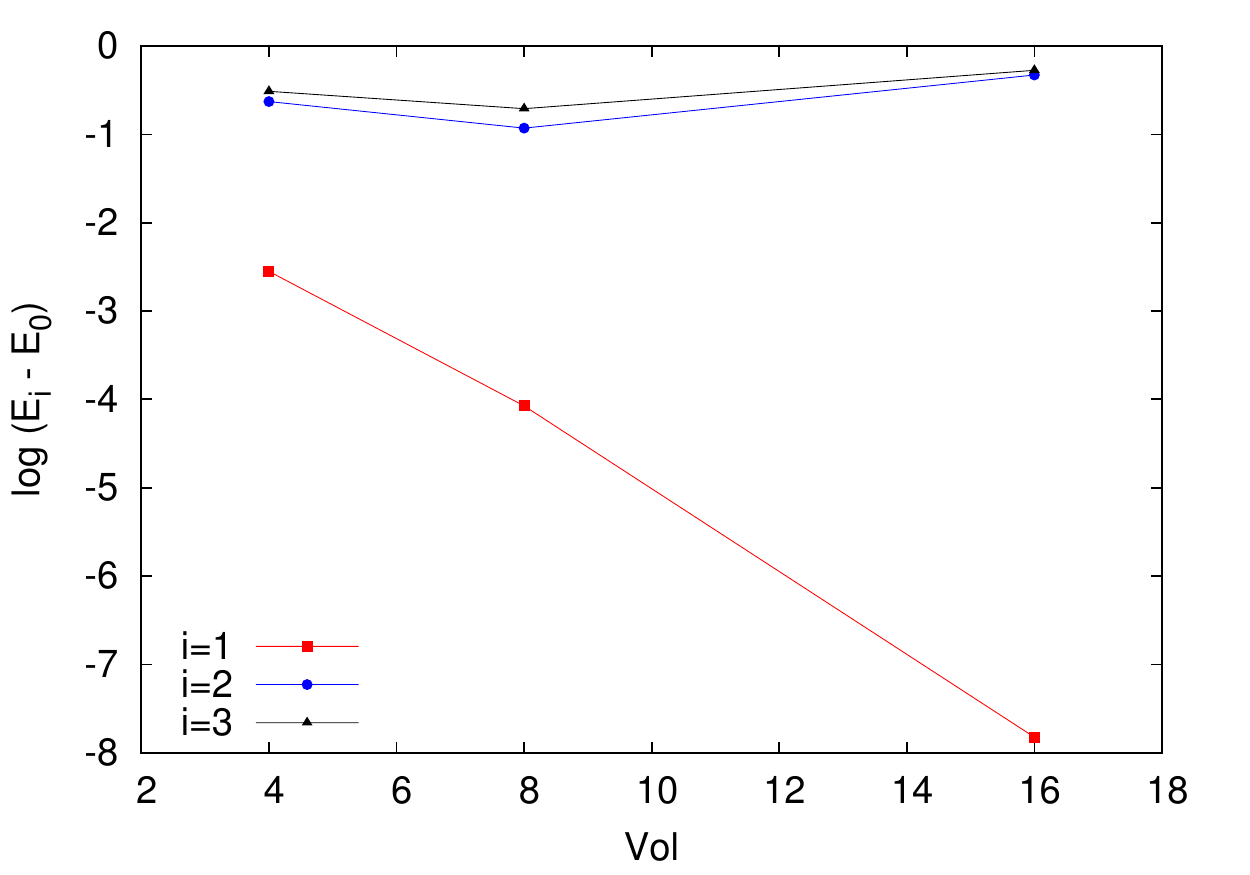}
\caption{{\it Energy gap of the pure $SO(3)$ QLM in $(2+1)$}-d. If the translation symmetry is spontaneously broken, then in the infinite volume the states with momenta $(0,0)$ and $(\pi,\pi)$ should be degenerate. The approach to this limit is exponential in the volume. Our results for $\log (E_1 - E_0)$ are consistent with this hypothesis. Note that the higher gaps $E_2 - E_0$, $E_3 - E_0$ are insensitive to the increase in volume. \label{fig:spect}}  
\end{center}
\end{figure}

Following our studies in the $(1+1)$-d case, it is not unreasonable to expect that certain lattice symmetries can be broken in the ground state. To understand this, we have studied the $(2+1)$-d model on small lattices by using exact diagonalization techniques. As explained in the previous section, the Gauss law can be analytically imposed to obtain two gauge invariant states per lattice site. We have studied system sizes up to $L^2 = 16$ using exact diagonalization to obtain the complete spectrum of the model. Note that since the effort increases exponentially with the volume, (much) larger systems are out of reach. The Monte Carlo approach, on the other hand, suffers from a sign problem in the chosen basis of the Hilbert space. 

However, from the exact diagonalization results we can already infer some interesting features of the ground state of the model. In particular, we observe a gap in the spectrum which exponentially closes with volume. Since the first excited state has momenta $(\pi,\pi)$ in contrast to momenta $(0,0)$ for the ground state, this is consistent with the hypothesis that the symmetry of translations by one lattice spacing is spontaneously broken. The result is shown in Fig \ref{fig:spect}.

\section{External charges in the pure $SO(3)$ gauge theory}
\label{dcon}

In addition to the broken translation symmetry, we note that the $SO(3)$ gauge theory has a trivial center. The confinement-deconfinement phase transition in a pure lattice gauge theory is related to the breaking of the center symmetry of the Polyakov loop, which is the order parameter. With the matter fields added, the Polyakov loop is no longer an order parameter, and one has to study the related Fredenhagen-Marcu order parameter~\cite{Marcu}. In addition, the potential between two static external charges is an extensively studied quantity in QCD~\cite{StrBr}, which we now study for the $SO(3)$ model. 

In the $SO(3)$ model, different charges can be considered: triplet (charge 1) and qunitet (charge 2). While the triplet charge can be realized in three different ways, there is only a single way to realize the quintet charge. To study the system in the presence of external charges, we impose the Gauss law everywhere except at the position of the charges, where one either imposes the triplet or the quintet Gauss law.     

As an example, we consider putting two triplet charges at a distance $x$. The charged states are chosen from the 9-dimensional subspace such that they belong to the $S^3_{\mathrm{tot}} = 1$ triplet:
\begin{equation}
\begin{split}
&| 1 \rangle = \frac{1}{\sqrt{2}} | \uparrow_1 \uparrow_3 \rangle 
 \left( |\uparrow_{2} \downarrow_{4} \rangle - | \downarrow_{2} \uparrow_{4} \rangle  \right), \\
&| 2 \rangle = \frac{1}{\sqrt{2}} | \uparrow_2 \uparrow_4 \rangle 
 \left( |\uparrow_{1} \downarrow_{3} \rangle - | \downarrow_{1} \uparrow_{3} \rangle  \right), \\
&| 3 \rangle =  \frac{1}{2}  | \uparrow_1 \uparrow_3 \rangle 
 \left( |\uparrow_{2} \downarrow_{4} \rangle + | \downarrow_{2} \uparrow_{4} \rangle  \right) \\
& ~~~~ -  \frac{1}{2}  | \uparrow_2 \uparrow_4 \rangle 
 \left( |\uparrow_{1} \downarrow_{3} \rangle + | \downarrow_{1} \uparrow_{3} \rangle  \right). 
\end{split}
\end{equation}

\begin{figure}[t!]
\begin{center}
\includegraphics[width = 0.65\columnwidth]{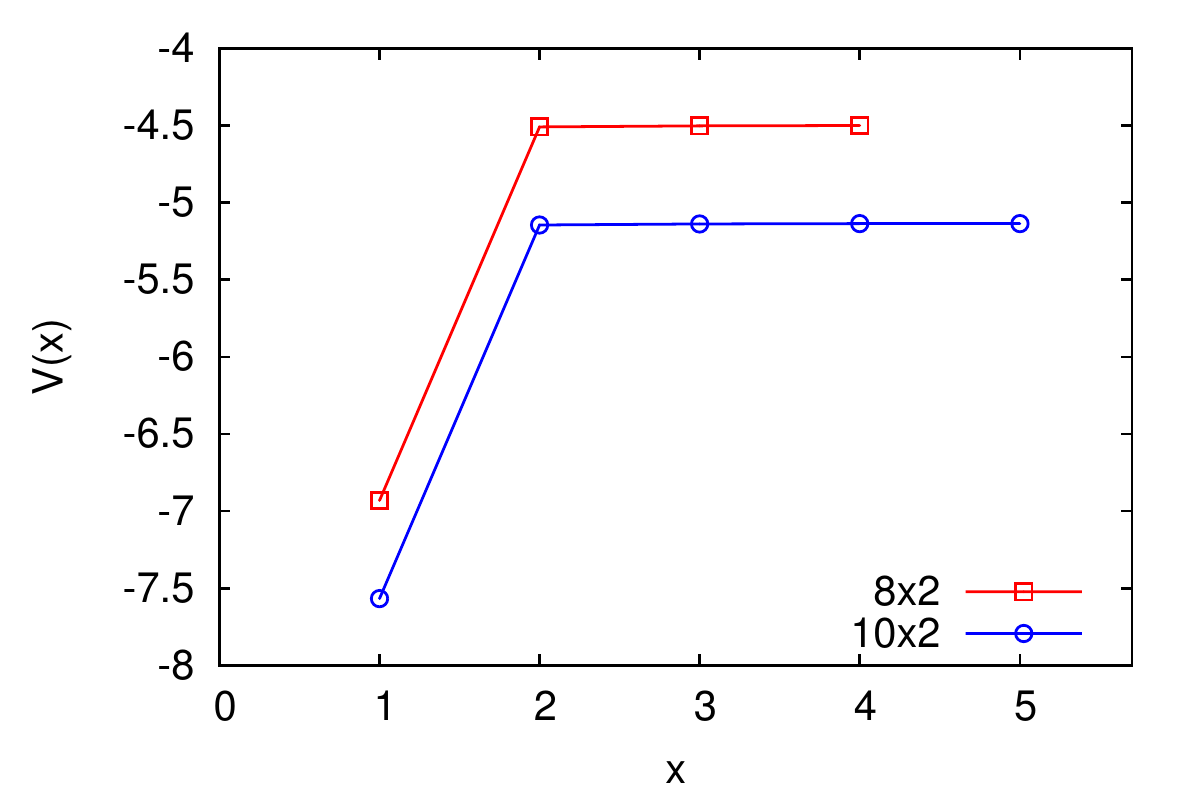} \\
\includegraphics[width = 0.65\columnwidth]{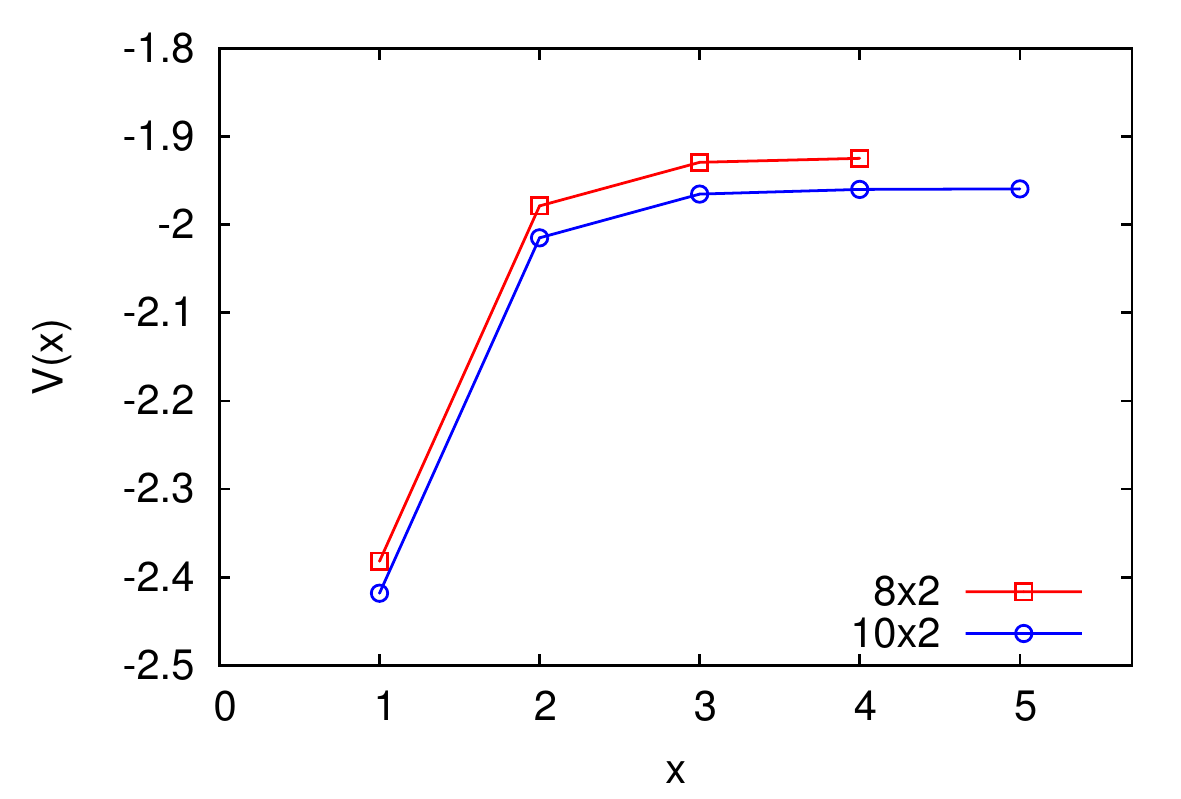}
\caption{{\it Potential with triplet (top) and quintet (bottom) charges.} The figures show the potential as a function of spatial separation of the triplet and quintet charges in a ladder system. In both cases the rapid screening of the external charges by the "gluons" becomes evident in the flatness of the potential as a function of the separation between the charges. \label{fig:charges}}  
\end{center}
\end{figure}

 Different possible realizations of the triplet charge are related by gauge transformations, and thus it is sufficient to consider a single realization of the triplet charge. We have studied the model with triplet and quintet charges on ladder systems as a function of different spatial separations. The results for the triplet and quintet charges are shown in Fig \ref{fig:charges}. In both cases, there are some finite volume effects, but the qualitative nature remains the same. As the charges are pulled apart, the "gluons" screen the charges almost immediately, and beyond two lattice spacings the individual charges are insensitive to the presence of the other charge. This behavior is different from that of QCD, where the potential has a linear rise due to the confining string joining the static charge-anti-charge, but becomes flat as soon the energy of the string matches the energy to create a dynamical quark-anti-quark pair from the vacuum. In QCD, the gluons are in the adjoint representation and hence cannot screen the static charges, which are in the fundamental representation. However, in our model, the "gluons" and the static external charges are in the same representation of $SO(3)$, and the "gluons" can screen the external charges. Thus, we note that while both QCD and the $SO(3)$ quantum link model are confined, the static charge-anti-charge potential behaves quantitatively very differently. 

These two features combined point to a rather exotic phase: confined, but with a broken translational invariance. In previous studies of the $U(1)$ quantum link model in $(2+1)$-d, different kinds of confined phases were observed which break translation symmetry {\color{blue} ~\cite{qlink1,qlink2}}. This study suggests the existence of other crystalline confined phases in non-Abelian quantum link models which, in addition, also screen external charges.
\end{document}